\documentclass[aps,prc,twocolumn,amssymb,tightenlines,showpacs]{revtex4}
\usepackage[dvips]{graphicx}  
\usepackage{dcolumn}          
\usepackage{bm}               
\begin{document}
\def\bea{\begin{eqnarray}}
\def\eea{\end{eqnarray}}
\def\be{\begin{equation}}

\def\ee{\end{equation}}
\def\rra{\right\rangle}
\def\lla{\left\langle}
\def\tv{\bm{\tau}}
\def\sv{\bm{\sigma}}
\def\rv{\bm{r}}

\title{Microscopic three-body forces and kaon condensation in cold
neutrino-trapped matter}

\author{ A.~Li$^{1}$, G.~F.~Burgio$^{2}$, U.~Lombardo$^{3}$, W.~Zuo$^{1}$}

\affiliation{$^1$ School of Physical Science and Technology,
Lanzhou University, and Institute of
Modern Physics, Chinese Academy of Sciences, Lanzhou 730000,
P.~R.~China
\\
$^2$ INFN Sezione di Catania, Via Santa Sofia 64, I-95123 Catania,
Italy\\
$^3$ INFN-LNS, Via Santa Sofia 44, I-95123 Catania,  and
Department of Physics and Astrophysics, Catania University, Via
Santa Sofia 64, I-95123, Italy}

\date{\today}

\begin{abstract}

We investigate the composition and the equation of state of the kaon
condensed phase in neutrino-free and neutrino-trapped star matter
within the framework of the Brueckner-Hartree-Fock approach with
three-body forces. We find that neutrino trapping shifts the onset
density of kaon condensation to a larger baryon density, and reduces
considerably the kaon abundance. As a consequence, when kaons are
allowed, the equation of state of neutrino-trapped star matter
becomes stiffer than the one of neutrino free matter. The effects of
different three-body forces are compared and discussed. Neutrino
trapping turns out to weaken the role played by the symmetry energy
in determining the composition of stellar matter, and thus reduces
the difference between the results obtained by using different
three-body forces.

\end{abstract}

\pacs{
 21.65.+f,  
 26.60.+c,  
 24.10.Cn,  
 97.60.Jd,  
 13.75.Cs   
     }

\maketitle

\section{Introduction}

Within a newly born neutron star or proto-neutron star (PNS)
neutrinos are temporarily trapped in a time scale of several tens
of seconds, which would undoubtedly contribute to the overall
chemical equilibrium, and may change the structure and composition
of the star. Neutrino trapping is expected to have a strong
influence on the equation of state (EOS) of $\beta$-stable neutron
star matter~\cite{Prakash:1997}, and, as a consequence, affects the
theoretical predictions of the maximum mass and the corresponding
radius. The EOS of the neutrino trapped stellar matter is
particularly important for the evolution of a newly born neutron
star and the mechanism of black hole formation. The effects
of neutrino trapping in stellar matter and its implications for
astrophysical phenomena were explored by several
authors~\cite{Prakash:1997,pons:1999,pons:2001,Vidana:2003,santos:2004,
panda:2004,nbbs:2005}. In Ref.\cite{Prakash:1997}, Prakash et al.
investigated systematically the structure and composition of
PNSs by using various theoretical models including the schematic
potential model ~\cite{Prakash:1988}, the
relativistic field approach~\cite{glendenning:1985} based on the
Walecka model~\cite{walecka:1974}, and the effective chiral
model~\cite{kaplan:1986}. It was argued that a black hole would be
most likely formed in a neutrino diffusion timescale of $\approx$ 10 s,
if the maximum mass of the hot neutrino-trapped star is about $1.5M_\odot$
in presence of negatively-charged hadrons. In
Ref.\cite{Vidana:2003,nbbs:2005}, the presence of neutrinos was shown to
delay the onset of hyperons, thus changing significantly the composition
of star matter, and making the EOS stiffer. In
Ref.~\cite{pons:2001}, the evolution of PNSs with quarks was
explored.

In the interior of a neutron star (NS) or PNS, the baryon density
could be as high as several times the value of the nuclear matter
saturation density, and thus the matter is expected to become
exotic. Among the possible exotic phases in dense nuclear matter,
the kaon condensation is a subject of great interest in nuclear
physics, hadronic physics and neutron star
physics~\cite{kaplan:1986,lee:1996,Li:1997,brown:1998,muto:2004}.
The presence of a kaon condensation may have important consequences
for determining
structure~\cite{thorsson:1994,pons:2000,pal:2000,banik:2001},
cooling rates~\cite{brown:1988,tatsumi:1988,fujii:1994,kubis:2003}
and evolution of neutron stars~\cite{pons:2001a}. The phase
transition from normal matter to the kaon condensed matter is also
expected to affect the transport properties and the glitch behavior
of pulsars~\cite{glen_schaf}. Other strange phases could also appear
in competition or coexisting with kaons; in particular, hyperons are
expected to occur at $2-3$ times the saturation density
\cite{Prakash:1997}, before the onset of kaon condensation.

In Ref.\cite{Zhou:2004}, the effects of different three-body forces
on the neutron star maximum mass were investigated within the
Brueckner-Hartree-Fock (BHF) approach. In Ref.\cite{zuo:2004}, the
kaon condensation in neutrino-free NS matter was studied in the same
theoretical approach. It was found that the composition of NS matter
is sensitively dependent on the nuclear symmetry energy and the
presence of kaons. In the present paper we shall extend the previous
work to the study of neutrino trapped matter at zero temperature, focussing
particularly on the interplay between the roles played by TBFs,
neutrino trapping and kaon condensation. Effects due to finite temperature,
which are important for the internal composition of protoneutron stars,
and their evolution \cite{pons:2000,pons:2001,tatsumi:1998,muto:2000}, will
be discussed within the  Brueckner-Hartree-Fock
approach, in a subsequent paper\cite{li:2006}. .

The present paper is organized as follows. In Sect.2 we review
briefly the theoretical models adopted in our calculations,
i.e., the BHF approach, and the chiral model for kaon-nucleon interaction.
In Sect.3 we discuss the stellar matter composition both with and without
neutrino trapping, and the resulting EOS.
Our numerical results are presented in Sect.4.
Finally, a summary is given in Sect.5.

\section{Theoretical models}\label{model}

\subsection{Brueckner-Bethe-Goldstone theory}

In the present work, we employ the Brueckner approach for
asymmetric nuclear matter~\cite{bombaci:1991,zuo:1999} to
calculate the baryonic contribution to the EOS of the stellar
matter. The starting point of the BHF approach is the interaction
$G$ matrix which satisfies the Brueckner-Bethe-Goldstone (BBG)
equation\cite{bombaci:1991,zuo:1999}
\begin{equation}
G(\rho,x_p;\omega)
= v_{NN} + v_{NN} \sum_{k_1 k_2}
 \frac { |k_1 k_2 \rangle Q \langle k_1 k_2| }
{\omega - \epsilon(k_1)-\epsilon(k_2)} G(\rho,x_p;\omega),
\label{eq:BG}
\end{equation}
where $\omega$ is the starting energy, and $x_p=\rho_p/\rho$ is the
proton fraction, being $\rho_p$, and $\rho$
the proton and the total baryon density, respectively.
$Q$ is the Pauli operator,
which prevents the two intermediate nucleons from being scattered
into the states below the Fermi sea, and $\epsilon(k)$ is the
single particle energy given by $\epsilon(k)
\equiv\epsilon(k;\rho) = \hbar^2k^2/(2m)  + U(k;\rho)$. The
single particle potential $U(k)$ is calculated from the real part
of the on-shell $G$-matrix and we adopt for it the so
called $continuous~ choice$~\cite{jeukenne:1976}, i.e.
for any momentum $k$ below and above the Fermi surface

\begin{equation}
U(k;\rho) = {\rm Re} \sum _{k'\leq k_F} \langle k k'|G[\rho; e(k)+e(k')]|k k'\rangle_a,
\label{eq:uk}
\end{equation}
\noindent
where the subscript ``{\it a}'' indicates
antisymmetrization of the matrix element. Due to the occurrence of
$U(k)$ in Eqs.~(\ref{eq:BG}) and (\ref{eq:uk}), the latter constitute a
coupled system of equations that has to be solved in a
self-consistent way. In the BHF approximation the energy per
nucleon is
\begin{equation}
{E \over{A}}  =
          {{3}\over{5}}{{k_F^2}\over {2m}}  + {{1}\over{2\rho}}
~ \sum_{k,k'\leq k_F} \langle k k'|G[\rho; e(k)+e(k')]|k k'\rangle_a.
\end{equation}
\noindent Adopting the continuous choice for the single-particle
potential, the two hole-line (BHF) truncation of the energy shift
turns out to be a good approximation for the nuclear EOS, since
the results in this scheme are quite close to those obtained by
including also the three hole-line contribution~\cite{song:1998}.
The basic input of the BBG equation is the realistic baryon-baryon
interaction, which is determined by reproducing the nucleon-nucleon
scattering phase shifts. The realistic nucleon-nucleon (NN)
interaction $v_{NN}$ adopted in the present calculation is the
Argonne $v_{18}$ two-body force~\cite{wiringa:1995}.
\begin{figure}
\includegraphics[totalheight=8.5cm,angle=270]{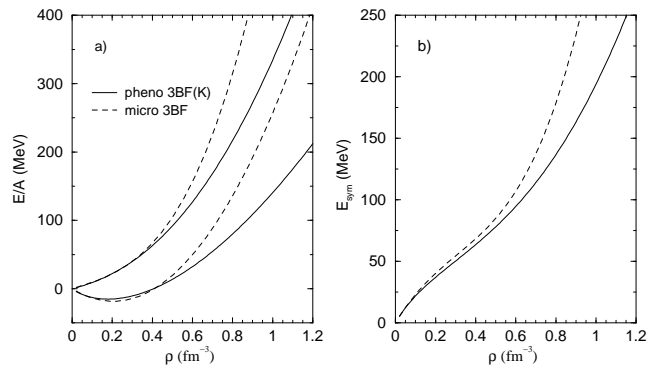}
\caption{The EoS is shown (panel (a)) for symmetric (lower curves) 
and pure neutron matter (upper
curves). The symmetry energy as a function of the nucleon density is 
displayed in panel (b).
See text for details.}
\label{f:esym}
\end{figure}
Since nonrelativistic calculations, based on purely two-body
interactions, fail to reproduce the correct saturation point of
symmetric nuclear matter, three-body forces (TBFs) among nucleons
are usually introduced \cite{baldo:1997}. They turn out to be needed
to correct this deficiency. In this work two models have been used,
i.e., the phenomenological Urbana model~\cite{pudliner:1995}, and a
microscopic TBF constructed from the meson-exchange current
approach~\cite{grange:1989}.
In both cases, the BBG approach is applicable only if the TBF is
reduced to an effective two body force, which has to be added to the
bare two body interaction. This is done  by averaging over the
position of the third particle, assuming that the probability of
having two particles at a given distance is given by the two-body
correlation function. The resulting potential is a density dependent
two-body force~\cite{baldo:1997}. The inclusion of the microscopic
TBF in the calculation for asymmetric nuclear matter has been
performed in Ref.~\cite{zuo:2000}.

In Fig.1, panel a), we show the equation of state for symmetric
nuclear matter (lower curves) and pure neutron matter (upper
curves). The solid lines represent the values obtained using the
Urbana phenomenological TBF, whereas the dashed lines are the
values obtained by adopting the microscopic TBF. We notice that
the results of the two adopted TBFs show a similar behavior up
to density $\rho \approx 0.4~fm^{-3}$, but they differ a lot in
the high density range. In particular, the microscopic TBF turns
out to be more repulsive than the Urbana model at high
densities, and the discrepancy between the two predictions becomes
increasingly large as the density increases.

Nucleonic TBFs play an essential role in determining the chemical
composition of a neutron star. In fact, as shown below, the baryon
chemical potentials and the symmetry energy strongly depend on the
choice of TBFs, for densities typical of those encountered in the
neutron star cores. The proton and neutron chemical potentials can
be derived from the energy density through the standard
thermodynamic relation
\begin{equation}
\begin{array}{lll}
\mu_{p,n} = (\partial \varepsilon / \partial \rho_{p,n})_{\rho_i} ,%
\end{array}
\end{equation}
with all remaining densities $\rho_i$ fixed.
In asymmetric nuclear matter, the calculation involves the knowledge
of the energy per particle and its partial derivatives with respect
to the total baryon density and proton fraction, i.e.
\begin{eqnarray}
\mu_{n}(\rho,x_{p})& = & E/A + \rho\frac{\partial E/A}{\partial\rho}
-x_{p}\frac{\partial E/A}{\partial x_{p}} \\
\mu_{p}(\rho,x_{p}) &= &E/A+\rho\frac{\partial E/A}{\partial\rho}
+(1-x_{p})\frac{\partial E/A}{\partial x_{p}}
    \label{mup:eps}
\end{eqnarray}
\noindent
Actually, the energy per particle $E/A$ may be
expanded quadratically in the proton concentration $x_p$ about its
value for symmetric matter ($x_p=\frac{1}{2}$):
\begin{equation}
\frac{E}{A}(\rho,x_p) = \frac{E}{A}(\rho,\frac{1}{2})
+{(1-2x_p)}^2E_{sym} + {\cal O}((1-2x_p)^4) \label{etot}
\end{equation}
Microscopic investigations~\cite{Prakash:1988,bombaci:1991,zuo:1999} have
shown that the expansion up to the quadratic term is a good
approximation and the higher-order contributions is negligible. As
a consequence, the difference of the neutron and proton chemical
potentials is determined by the symmetry energy in an explicit
way
\begin{equation}
\mu_n - \mu_p=4(1-2x_p)E_{sym}(\rho)
\end{equation}
which implies that the symmetry energy plays a decisive role in
predicting the chemically equilibrated compositions of PNSs and NSs.

In Fig.1, panel b), we show the symmetry energy calculated in the
BHF approach as a function of the nucleon density. Its value at
the normal nuclear matter density has been well determined to be
$30\pm4$ MeV. Actually, the calculations performed with either the
Urbana phenomenological TBF (solid line) or the microscopic TBF
(dashed line) reproduce correctly this value. However, the high
density behavior suffers a large uncertainty, and this reflects in
the theoretical predictions of the neutron star composition. Moreover,
we see that the symmetry energy predicted by using
the microscopic TBF rises much faster at high densities than the one
obtained by adopting the phenomenological TBF.

As far as leptons are concerning, the energies and chemical
potentials are obtained by solving the free Fermi gas model at zero
temperature \cite{Shapiro:1983}. Once the baryonic and leptonic
chemical potentials are known, one can proceed to calculate the
composition of the $\beta$-stable stellar matter, and then the total
energy per particle $(E/A)_{tot}$ and pressure $P$ through the usual
thermodynamical relation
\begin{equation}
P = \rho^2 \frac{d (E/A)_{tot}}{d\rho}
\end{equation}

\subsection{Kaon condensation}

In the present paper, we use the effective chiral theory proposed by
Kaplan and Nelson~\cite{kaplan:1986} and extensively investigated
afterwards~\cite{brown:1992}. In this model, the $SU(3)\times SU(3)$
chiral Lagrange density is expressed as
\begin{eqnarray}
{\mathcal  L_\chi} & = & \frac{f^2}{4}{\mathrm Tr} \partial_\mu U
\partial^\mu
U^\dagger +{\mathrm Tr}\bar{B}(i \gamma^\mu D_\mu -  m_B) B  \nonumber \\
& & + \; F\, {\mathrm Tr} \bar{B}\gamma^\mu \gamma_5 [ {\mathcal
A}_\mu,B] + \; D\, {\mathrm Tr} \bar{B}\gamma^\mu \gamma_5 \{
{\mathcal A}_\mu,B\}
\label{lan-KN} \nonumber \\
& & + c {\mathrm Tr} {\mathcal M} (U + U^\dagger) + a_1 {\mathrm Tr}
\bar{B}(\xi {\mathcal M} \xi + \xi^\dagger {\mathcal M}
\xi^\dagger) B \nonumber \\
& & + a_2 {\mathrm Tr} \bar{B}B(\xi {\mathcal M} \xi + \xi^\dagger
{\mathcal M} \xi^\dagger) \\ 
& & + a_3 {\mathrm Tr} \bar{B}B\; {\mathrm
Tr}(\xi {\mathcal M} \xi + \xi^\dagger {\mathcal M} \xi^\dagger).
\end{eqnarray}
The first four terms conserve the chiral symmetry, and the other
terms break the chiral symmetry. Parameters in the symmetric part
are $f=93~ \mathrm{MeV}$, the pion decay constant, $D=0.81$, and
$F=0.44$. The breaking strength is determined by the parameters
$a_1, a_2, a_3$, and $c$, as well as by the quark mass matrix
${\mathcal M}$.  We adopt $a_1m_s=-67$ MeV and $a_2m_s=134$ MeV, as
determined from the baryon mass splittings~\cite{politzer:1991}. The
constant $c$ and the bare kaon mass are related by the
Gell-Mann-Oakes-Renner relation $m_K^2 = 2cm_s/f^2$. The parameter
$a_3m_s$ had remained largely uncertain for many years, due to our
poor knowledge of the strangeness content of the proton and the
kaon-nucleon sigma term
$\Sigma_{KN}$~\cite{lee:1996,kaplan:1986,politzer:1991,donoghue:1985}.
Fortunately, the large ambiguity in this parameter has been settled
recently by Dong, Laga\"{e} and Liu~\cite{dong:1996} with small
error based on the lattice calculations. In the present calculations
we adopt values of $a_3m_s$ equal to $-310,
-222, {\rm and} -134\ \mathrm{MeV}$, as done in
Refs.\cite{thorsson:1994,kubis:2003}, in order to investigate the
sensitivity of our results to the variation of the proton
strangeness content. Our adopted central value $a_3m_s =-222$ MeV is
very close to the value extracted from the lattice gauge
calculations in Ref.~\cite{dong:1996}, which is $-231$ MeV with
error of less than $4\%$. The value of $a_3m_s$ is essential in
determining the onset density for kaon condensation, since it
provides the attractive component of the kaon-nucleon interaction.

In principle, the chiral Lagrangian should give the EoS of baryons
and mesons, but so far  only the kaon-nucleon interaction has been
extracted from it. Therefore, we use the chiral Lagrangian only to
extract the kaon-nucleon part of the interactions, and take the EoS
for nuclear matter from the Brueckner many-body theory. This means
that the energy density is a sum of three contributions: the
kaon-nucleon interaction energy density (derived from $\mathcal
L_\chi$), the nucleon-nucleon energy density, and the energy density
for leptons ($e,\mu$):
\begin{equation}
\epsilon = \epsilon_{KN} + \epsilon_{NN} + \epsilon_{lep}.
\end{equation}
From eq.(\ref{etot}), we get the nucleon-nucleon contribution as
\begin{equation}
\epsilon_{NN}(\rho, x_p) = \rho \frac{E}{A}(\rho, x_p),
\end{equation}
whereas the energy density for leptons is the one for a free Fermi
gas at zero temperature, and can be found in textbooks
\cite{Shapiro:1983}.

Following exactly the standard prescription in
Ref.~\cite{thorsson:1994}, we can get the kaon-nucleon energy
density of the kaon condensed matter from applying the Baym
theorem~\cite{baym}, i.e.
\begin{eqnarray}
\epsilon_{KN} & = & \frac{f^2}{2} \mu_K^2\sin^2\theta +
2m^2_Kf^2\sin^2\frac{\theta}{2} \\
& + &
\rho(2a_1x_p+2a_2+4a_3)m_s\sin^2\frac{\theta}{2}
\end{eqnarray}
where $\rho$ and $x_p$ denote the nucleon number density and the
proton fraction, respectively, and $\theta$ is the amplitude of the
condensation.
We remind the reader that in this work we neglect the effects of
$\bar{K}^0$ condensation, which could play a significant role at the
high densities typical of neutron star cores, as it has been shown
in Ref.\cite{pal:2000,banik:2001}.

\section{Equation of state of cold catalyzed stellar matter}

For stars in which the strongly interacting particles are only
baryons, the chemical composition is determined by the
requirements of charge neutrality and equilibrium under the weak
processes

\begin{equation}
 B_1 \rightarrow B_2 + l + {\overline \nu}_l,
  \:\:\:\:\:  B_2 + l \rightarrow B_1 + \nu_l
    \label{weak:eps}
\end{equation}
where $B_1$ and $B_2$ are baryons, $l$ denotes a lepton, either an
electron or a muon, and $\nu_l$($\bar\nu_l$) a (anti-)neutrino. Under
the condition of neutrino escape, these two requirements imply
that the relations
\begin{equation}
 \sum_{i}q_{i}x_{i}+\sum_{l}q_{l}x_{l}=0
    \label{neutral:eps}
\end{equation}
\noindent
\begin{equation}
 \mu_{i}=b_{i}\mu_{n}-q_{i}\mu_{l}
    \label{mufre:eps}
\end{equation}
are satisfied. In the above expression, $x_i=\rho_i/\rho_B$
represents the baryon fraction for the species $i$, and $\rho_B$
the baryon density. The neutron chemical potential is denoted by
$\mu_n$, whereas $\mu_i$ refers to the chemical potential of the
baryon species $i$, $b_{i}$ to its baryon number and $q_{i}$ to
its electric charge. The same notation holds true for those
quantities with subscript $l$, i.e. leptons. Under condition when
the neutrinos are trapped in the system, the beta equilibrium
condition (\ref{mufre:eps}) is altered to
\begin{equation}
 \mu_i=b_i \mu_n - q_i(\mu_l-\mu_{\nu_l})
    \label{nitrap:eps}
\end{equation}
where $\mu_{\nu_l}$ is the chemical potential of the neutrino $\nu_{l}$.
\noindent

Because of the trapping, the numbers of leptons per baryon of each
flavor of neutrino $l=e,\mu$,
\begin{equation}
 Y_{L_{l}}=x_{l}+x_{\nu_{l}},
    \label{lepfrac:eps}
\end{equation}
are conserved on dynamical time scales. Gravitational collapse
calculations of the white-dwarf core of massive stars indicate that at
the onset of trapping, the electron lepton number
$Y_{L_{e}}=x_{e}+x_{\nu_{e}}\simeq 0.4$, the precise value depending
on the efficiency of electron capture reactions during the initial
collapse stage. Also, because no muons are present when neutrinos
become trapped, the constraint $Y_{L_{\mu}}=x_{\mu}+x_{\nu_{\mu}}=0$
can be imposed. We fix $Y_{L_{l}}$ at these values in our calculations
for neutrino trapped matter. \par
\begin{figure}
\includegraphics[totalheight=8.5cm,angle=270]{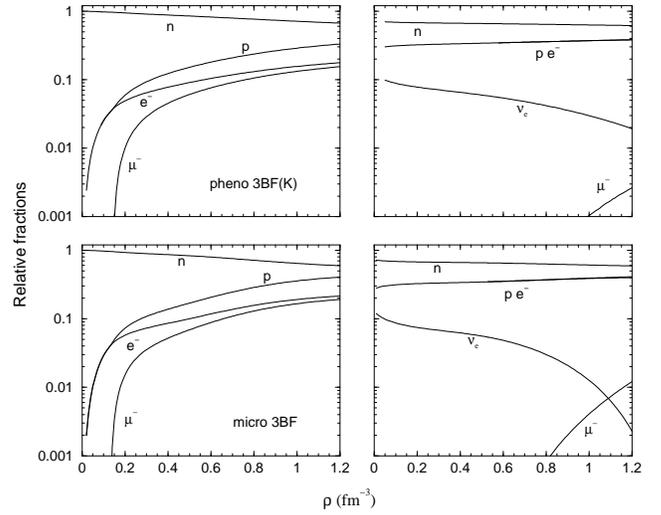}
\caption{The particle populations are shown as a function of the nucleon 
density for neutrino free
(left hand panels) and neutrino trapped matter (right hand panels). In the 
upper (lower) panels
results are displayed for the case when the phenomenological Urbana 
(microscopic) TBF is used.}
\label{f:comp}
\end{figure}
\noindent
\begin{figure}
\includegraphics[totalheight=8.5cm,angle=270]{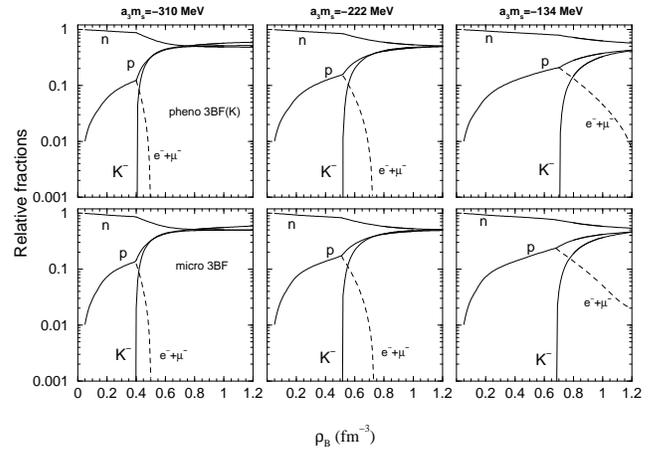}
\caption{The particle populations are shown as a function of the baryon 
density for neutron
star matter with kaons, for three different values of $a_3m_s$. In the upper 
(lower) panels
results are displayed for the case when the phenomenological Urbana 
(microscopic) TBF is used.}
\label{f:compk}
\end{figure}

In the neutron star matter with kaons the chemical equilibrium can
be reached through the following reactions
\begin{equation} n
\leftrightarrow p + l + \nu_l, ~~~ n \leftrightarrow  p + K^-
,\label{beta} ~~~ l \leftrightarrow  K^- + \nu_l ,\nonumber
\end{equation}
where $l$ denotes leptons, i.e., $l = e,\mu$. One can determine
the ground state by minimizing the total energy density with
respect to the condensate amplitude $\theta$ keeping all densities
fixed. This minimization together with the chemical equilibrium
and charge neutrality conditions leads to the following three
coupled equations~\cite{thorsson:1994,kubis:2003}\noindent
\begin{eqnarray} 
\cos\theta &=&  \frac{1}{f^2 \mu^2} [m_K^2 f^2 +
\frac{1}{2}u\rho_0(2a_1 x_p\!+\!2a_2 \!+\!4a_3)m_s~~~~ \\
  & - &  \frac{1}{2}\mu u\rho_0 (1\!+\!x_p)],
\label{th-min}
\end{eqnarray}
and
\begin{eqnarray} 
\mu &\equiv&
\mu_e - \mu_{\nu_e} =\mu_K = \mu_n-\mu_p \\
&=& 4(1-2x_p)S(u)\sec^2\frac{\theta}{2} - 
2a_1 m_s \tan^2\frac{\theta}{2} \,, \label{betatheta}
\end{eqnarray}
\begin{eqnarray}
 f^2\mu
 \sin^2\theta + u\rho_0 (1+x_p)\sin^2\frac{\theta}{2} - x_pu\rho_0 & & \\
+  \frac{ {\mu_e}^3 }{ 3\pi^2 }
 + \eta(|\mu|-m_\mu) \frac{ (\mu_\mu^2-m_\mu^2)^{3/2}}{ 3\pi^2 } & =& 0,
\nonumber \\
\label{cntheta}
\end{eqnarray}

where $u\equiv\rho_B/\rho_0$ is the baryon number density in units of
$\rho_0$. The last two equations are from the chemical equilibrium
and charge neutrality conditions, respectively.
For the neutrino-free case, the above equations recover the ones
given in Ref.\cite{zuo:2004}.
The EOS and the
composition of the kaon condensed phase in the chemically
equilibrated neutron star matter can be obtained by solving the
coupled equations (\ref{th-min}),
 (\ref{betatheta}), and (\ref{cntheta}). The critical density for
kaon condensation is determined as the point above which a
real solution for the coupled equations can be found.

The stable configurations of a neutron star can be obtained
 from the well known hydrostatic equilibrium equations
  of Tolman, Oppenheimer and
  Volkoff~\cite{Tolman:1939,Oppenheimer:1939,Shapiro:1983}
 for the pressure $P$ and the enclosed mass
$m$
\begin{equation}
 \frac{dP(r)}{dr}=-\frac{Gm(r)\mathcal{E}(r)}{r^{2}}
 \frac{\Big[1+\frac{P(r)}{\mathcal{E}(r)}\Big]
 \Big[1+\frac{4\pi r^{3}P(r)}{m(r)}\Big]}
 {1-\frac{2Gm(r)}{r}},
    \label{tov1:eps}
\end{equation}
\begin{equation}
\frac{dm(r)}{dr}=4\pi r^{2}\mathcal{E}(r),
    \label{tov2:eps}
\end{equation}
once the equation of state $P(\mathcal E)$ is specified, being
$\mathcal E$ the total energy density ($G$ is the gravitational
constant). For a chosen central value of the energy density, the
numerical integration of Eqs.(\ref{tov1:eps}, \ref{tov2:eps})
provides the mass-radius relation. For the description of the NS's
crust, we have joined the hadronic equations of state above
described with the ones by Negele and Vautherin~\cite{negele:1973}
in the medium-density regime ($0.001~fm^{-3}<\rho<0.08~fm^{-3}$),
and the ones by Feynman-Metropolis-Teller~\cite{feynman:1949} and
Baym-Pethick-Sutherland~\cite{baym:1971} for the outer crust
($\rho<0.001~fm^{-3}$).

\begin{figure}
\includegraphics[totalheight=8.5cm,angle=270]{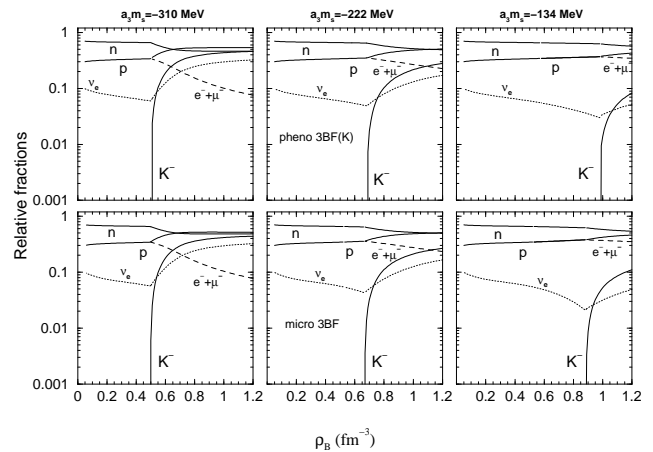}
\caption{Same as Fig.~\ref{f:compk}, but for neutrino trapped matter.}
\label{f:compk_nu}
\end{figure}

\section{Results and discussion}

In Fig.~\ref{f:comp} we present the composition of the
$\beta$-equilibrated neutrino-free (left-hand panels)
and neutrino-trapped $npe\mu$ matter (right-hand panels).
The upper panels display results obtained using the Argonne $v_{18}$
two-body potential supplemented by the phenomenological
Urbana TBF, whereas in the lower panels we show the populations
obtained when the microscopic TBFs are used.
In the neutrino-free case, we notice that a stiffer symmetry energy produces a
larger proton concentration in the region $\rho \geq 0.6~fm^{-3}$,
where the proton fraction $x_p \approx E_{sym}^3$.
In neutrino-trapped matter,
the lepton concentration becomes sizeably higher because
the electron chemical potential must keep at a high value, in order to
fulfill the $\beta$-equilibrium condition.
This leads to a larger proton fraction as compared to the neutrino-free
case, because of the charge neutrality. Therefore, neutrino-trapped matter
is more symmetric than the neutrino-free one.

We also notice that the differences in proton populations, due to
the different TBFs adopted, are smoothed out when neutrinos are
trapped. This is readily understood as follows. In neutrino-free
matter, the electron chemical potential is determined mainly by the
density dependence of nuclear symmetry energy. When neutrinos are
trapped, they contribute to the chemical equilibrium, and, as a
consequence, the electron chemical potential is determined
simultaneously by the neutrino trapping effect and by the symmetry
energy. Because of the lepton number conservation, the major effect
of trapping is to keep the electron concentration high, and this
reduces considerably the role played by the symmetry energy in
determining the star composition.

Let us now discuss the case when kaons are present in $\beta$ stable
and neutrally charged matter. In Fig.~\ref{f:compk} we show the
particle fractions for neutrino free star matter, using
alternatively the phenomenological Urbana TBF (upper panels) and the
microscopic one (lower panels). We have chosen three different
values of the parameter $a_3m_s$, respectively $a_3m_s= -310,
-222,{\rm and}-134~{\rm MeV}$. We observe that the kaon
threshold depends sensitively on the attraction term of the
kaon-nucleon interaction. In the extreme case of $a_3m_s=-310$ MeV,
the predicted kaon onset density is the lowest, of the order of $
0.4~ {\rm fm}^{-3}$, almost independently on the adopted nucleonic
TBF. This is due to the fact that, up to densities of the order of $
0.6~{\rm fm}^{-3}$, both kinds of TBF produce a very similar
symmetry energy. In addition, the role played by the kaon-nucleon
interaction becomes more predominant over that by the symmetry
energy for a stronger kaon-nucleon interaction term. For larger
densities, the microscopic TBF gives a higher symmetry energy, and
this leads to a slightly different kaon threshold when a weak
attractive term is adopted. We also notice a dramatic decrease of
the lepton population and increase of the proton concentration as a
consequence of the kaon appearance. This is due to the charge
neutrality condition, which has to be always fulfilled. We notice
that the threshold densities for kaon condensation may be delayed by
the onset of hyperons, as found by other authors
\cite{banik:2001,knor:1995,schaf:1996}.

In Fig.\ref{f:compk_nu} the particle population is plotted as a
function of the baryon density for the neutrino trapped case. We
notice that, because of trapped neutrinos, the onset of kaon
condensation is shifted to larger densities than in neutrino free
matter. Even in
that case, both TBF's produce very similar results, except for the
less attractive case, i.e. $a_3m_s=-134~{\rm MeV}$, where the onset
for kaon condensation takes place at a lower density for the
microscopic TBF. This is due to the fact that, at high density, the
symmetry energy is larger, and this allows for a kaon condensed
phase at baryon density lower than in the case with the Urbana TBF.
As compared to the neutrino free case, the kaon abundance in the
condensed phase is much smaller than in neutrino trapped matter, due
to the larger electron population. In contrast to the normal phase,
in the condensed phase the number of neutrinos rises up as the
baryon density increases, and this is because the presence of kaons
leads to a decrease of the electron population as a function of the
baryon density. We notice that the values of the threshold densities
for $K^-$ condensation both in neutrino-free and neutrino-trapped
matter are in very good agreement with those found in
ref.\cite{Prakash:1997}. However,
the threshold densities are dependent both on the value of the
temperature, and the attraction term of the kaon-nucleon
interaction \cite{Prakash:1997,pons:2001}.
\begin{figure}
\includegraphics[totalheight=8.5cm,angle=270]{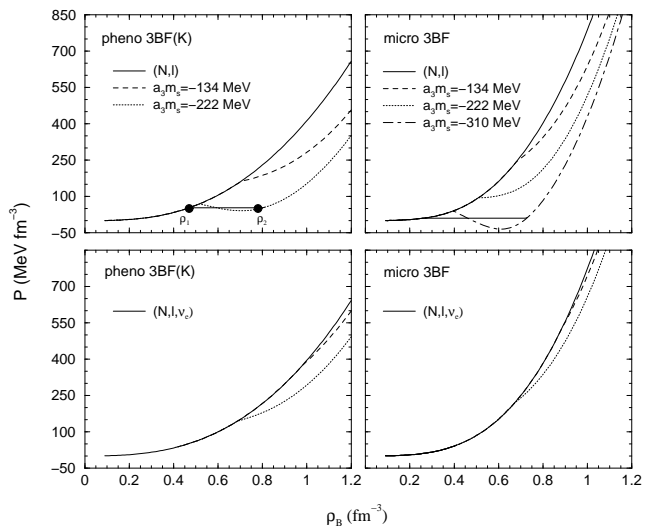}
\caption{The pressure for beta-equilibrated neutrino free (upper panels) and 
neutrino trapped matter
(lower panels) is shown as a function of the baryon density, for 
phenomenological (left-hand side)
and microscopic TBF's (right-hand side). The thin solid lines represent 
the Maxwell construction. }
\label{f:press_nu}
\end{figure}

Once the relative particle concentrations are known, we can
calculate the equation of state. This is shown in
Fig.\ref{f:press_nu}, both for the neutrino free case (upper
panels), and the neutrino trapped one (lower panels). In the case
without kaons, the equation of state of neutrino trapped matter is
slightly softer than the neutrino free one, because the loss in
energy due to the reduction of the neutron-proton asymmetry as a
consequence of the increase of the proton abundance exceeds the gain
from the presence of neutrinos~\cite{Prakash:1997}. We observe that
the kaon condensation produces a general softening of the equation
of state with respect to the purely nucleonic case. The degree of
softening depends on the value of the parameter $a_3m_s$, i.e., the
larger the value of $|a_3m_s|$ the stronger the softening is. In the
case with kaon condensate, neutrino trapping produces a stiffer
equation of state due to the higher onset density of kaons and
smaller kaon abundance, as shown in Fig.\ref{f:compk_nu}. It is
evident from the figure that the softening of the equation of state
due to kaon condensate is much less pronounced in neutrino trapped
matter than in neutrino free matter. This may lead a newly-formed,
hot protoneutron star to metastability, i.e. a delayed collapse
while cooling down, as discussed in
ref.\cite{Prakash:1997,pons:2000}.
\begin{figure}
\includegraphics[totalheight=8.5cm,angle=270]{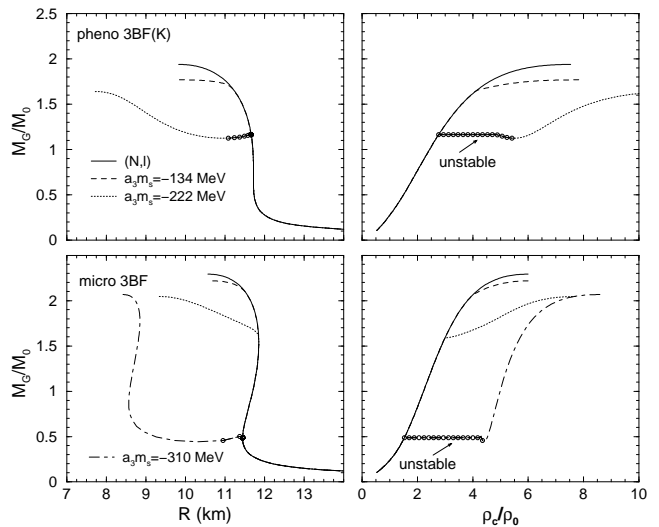}
\caption{The gravitational mass is shown as a function of the radius 
(left-hand side) and the
normalized central density (right hand side), for the phenomenological 
(upper panels) and microscopic TBF's
(lower panels) in neutrino free matter. Circles represent unstable 
configurations.}
\label{f:mass_rho_free}
\end{figure}

In some cases, the onset of a kaon condensed phase produces a
negative compressibility in the equation of state. Following Migdal
\cite{mi:1979}, we have performed a Maxwell construction to maintain
a positive compressibility. We should note that the Maxwell
construction is valid for substances with only one independent
component. Neutron star matter has two independent components (the
baryon and electric charge). Therefore, the Maxwell construction
cannot satisfy Gibbs' criteria that pressure, temperature and all
chemical potentials be common to both phases in equilibrium. A novel
treatment of the phase transition in $\beta$ stable matter has been
proposed by Glendenning\cite{glen:1992} for the hadron-quark phase
transition, and has been subsequently extended to the study
of the kaon condensed phase, treated as a first order phase
transition in neutron star matter \cite{glen_schaf}. The Gibbs
construction has strong consequences on the mechanical stability of
neutron stars, because the pressure is a monotonically increasing
function of the density. On the contrary, there is a mechanical
instability for the Maxwell case  that is initiated by the central
densities for which the pressure remains constant. Such an unstable
region is absent when the phase transition is treated using Gibbs'
conditions. It turns out that the value of the maximum mass
of neutron stars with kaon condensation is only slightly affected by
the Gibbs construction\cite{glen_schaf}, which is still
affected by many theoretical uncertainties \cite{maru:2006}.
Therefore we limit ourselves to the Maxwell construction.
In Fig.5, the region of
constant pressure (thin solid line) is comprised between two values
of the baryon density, denoted by $\rho_1$ and $\rho_2$. If the
magnitude of $|a_3m_s|$ is large, it may happen that
$\rho_1<\rho_0$. As a rule, we consider this solution as
unrealistic, and we do not show the corresponding curve in
Fig.\ref{f:press_nu}. This is the case when $a_3m_s=-310~\rm MeV$
and neutrinos are trapped, that is to say that the kaon condensation
requires a very stiff symmetry energy if the kaon-nucleon
interaction is strongly attractive.

Once the EOS has been determined, the TOV equations can be solved.
The resulting gravitational mass is plotted for neutrino free matter
in Fig.\ref{f:mass_rho_free}, as a function of both radius and
central density, normalized with respect to the saturation value. In
particular, microscopic TBF's produce stiffer equations of state,
and therefore give rise to larger values of the maximum mass. We
observe a decrease of the maximum mass when kaons appear, the exact
value being dependent on the chosen TBF's and the parameter
$a_3m_s$. In the microscopic case, the inclusion of kaons decreases
the maximum mass configuration down to about 2 $M_\odot$. On the
other hand, phenomenological TBF's produce softer equations of state
and, consequently, smaller neutron stars, whose maximum mass is
about 1.7 $M_\odot$. In Fig.\ref{f:mass_rho_free} circles represent
unstable configurations, originating from the Maxwell construction.
Those configurations turn out to be stable if the Gibbs
construction \cite{glen:1992,glen_schaf} is applied to the transition from
purely nuclear matter to the kaon condensed phase.
\begin{figure}
\includegraphics[totalheight=8.5cm,angle=270]{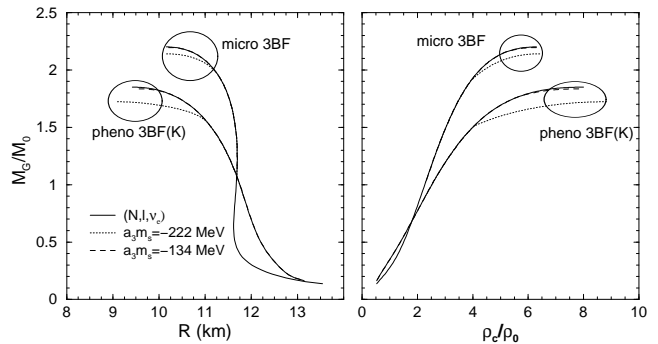}
\caption{The gravitational mass is shown as a function of the radius 
(left-hand side) and the
normalized central density (right hand side), for the phenomenological 
(lower curves)
and microscopic TBF's (upper curves) in neutrino trapped matter.}
\label{f:mass_rho_trap}
\end{figure}

In the purely nucleonic case, neutrino trapping generally produces
a softer equation of state, because beta-stable matter turns out
to be more symmetric in neutrons and protons. As a consequence,
the maximum mass becomes smaller.
This is displayed in Fig.\ref{f:mass_rho_trap}.
However, the appearance of kaons
changes this general picture and the equation of state becomes
stiffer. This is due to the fact that the $K^-$ onset depends on
the lepton chemical potential, i.e., $\mu_e-\mu_{\nu_e}$, which
stays at larger values in neutrino-trapped matter than in the
neutrino-free case, thus delaying the appearance of $K^-$ to
higher baryon density. The resulting maximum mass increases, as shown in
Fig.\ref{f:mass_rho_trap}. If phenomenological TBF's are used, the
value of the maximum mass stabilizes around 1.8 $M_\odot$, whereas
microscopic TBF's give rise to protoneutron stars with mass
slightly larger than 2 $M_\odot$. As compared to the neutrino free
case (Fig.\ref{f:mass_rho_free}), the reduction of the maximum
mass due to the presence of kaons is smaller in the neutrino
trapped case.

\section{Summary and conclusions}

In summary, we have investigated the properties of kaon condensed
neutrino-free and neutrino-trapped matter at zero temperature
including the composition, the equation of
state, and the radius-mass relation, in the framework of the BHF
approach with both microscopic and phenomenological three-body
forces. In particular, we have discussed the interplay among the
effects of the different TBFs, kaon condensate, and neutrino
trapping. It is found that the effect of the microscopic TBF
results in larger values of symmetry energy at high densities
($\rho_B\ge 0.4~{\rm fm}^{-3}$ ) than that of the phenomenological
one. This gives rise to a higher proton population in NS matter
without neutrinos and kaons, thus allowing for fast cooling through
the direct URCA process. The contribution of trapped neutrinos
makes the electron concentration keep high in $\beta$-equilibrated
star matter and weakens the role played by the symmetry energy on
the predicted composition of the star. As a result, the
composition of the star matter becomes less sensitive to the two
different three-body forces considered when neutrinos are trapped.
In the star matter without kaons, neutrino trapping leads to a
softening of the equation of state. However when kaons are
allowed, neutrino trapping makes the equation of state stiffer as
compared to the neutrino free star matter, since the presence of
trapped neutrinos shifts the onset of kaons to higher densities
and reduces the kaon abundance.

In both cases with and without kaons, the microscopic TBF leads to
larger values of the maximum star masses than the
phenomenological one. If only nucleons and leptons are allowed,
the effect of neutrino trapping is to reduce the maximum star
mass; while when kaons are allowed, trapping instead produces
larger stars. As expected, the effect of kaon condensate is
generally to soften the equation of state and reduce the predicted
maximum star mass. Neutrino trapping weakens the role of kaons by
delaying their onset densities to higher values and by reducing
the kaon abundance. As a result, the reduction of the maximum mass
due to kaon condensate is less in the neutrino trapped case than
in the neutrino free case.

This general scenario could change if we include hyperons in our
calculations. Several studies have been performed on hyperonic
matter within the BHF approach \cite{bbs,vid2000}, showing that
$\Sigma^-$ appear already at a very low value of the baryon density,
i.e., $\rho_B \approx (2-3) \rho_0$. Therefore, hyperon onset might
be a process in competition with kaon condensation
\cite{Prakash:1997}, as found in relativistic mean-field approaches
\cite{knor:1995,ell:1995,schaf:1996}. We will present a more detailed
study in a forthcoming publication.

\section{Acknowledgments}

One of us (W. Zuo) acknowledges the warm hospitality he received
at LNS-INFN, Catania where this work started. The work was done
within the Asia-Link project (CN/ASIA-LINK/008(94791)) of the
European Commission. 
The work of A. Li and W. Zuo was supported in part by the National Natural 
Science Foundation of China (10575119, 10235030), the Knowledge Innovative 
Project of CAS (KJCX2-SW-N02), the Major Prophase Research Project of 
Fundamental Research of the Ministry of Science and Technology of China
(2002CCB00200), the Chinese Major State Basic Research Development
Program (G2000077400), and the Knowledge Innovative Project of CAS
(KJCX2-SW-N02).


\begin{thebibliography}{99}

\bibitem{Prakash:1997}M. Prakash, I. Bombaci, M. Prakash, P. J. Ellis,
 R. Knorren, and J. M. Lattimer, Phys. Rep. {\bf 280}, 1 (1997)
  and references therein.

\bibitem{pons:1999}  J. A. Pons, S. Reddy, M. Prakash, J.
M. Lattimer, and J.A. Miralles, Astrophys. J. {\bf 513}, 780
(1999).

\bibitem{pons:2001} J. A. Pons, A. W. Steiner, M. Prakash, and J.
M. Lattimer, Phys. Rev. Lett. {\bf 86}, 5223 (2001).

\bibitem{Vidana:2003} I. Vida\~{n}a, I. Bombaci, A. Polls,
 and A. Ramos, Astronomy \& Astrophysics {\bf 399}, 687 (2003);
Nucl. Phys. {\bf A719}, 173c (2003).

\bibitem{santos:2004} A. M. S. Santos, and D. P. Menezes,
 Phys. Rev. {\bf C69}, 045803 (2004).

\bibitem{panda:2004} P. K. Panda, D. P. Menezes, and C.
Providencia, Phys. Rev. {\bf C69}, 025207 (2005).

\bibitem{nbbs:2005} O. E. Nicotra, M. Baldo, G. F. Burgio, and H.-J. Schulze,
Astronomy $\&$  Astrophysics {\bf 451}, 213 (2006).

\bibitem{Prakash:1988}M. Prakash, T. L. Ainsworth, and J. M.
Lattimer, Phys. Rev. Lett. {\bf 61}, 2518 (1988).

\bibitem{glendenning:1985} N. K. Glendenning, Astrophys. J.
 {\bf 293}, 470 (1985).

\bibitem{walecka:1974} J. D. Walecka,
Ann. Phys. (N.Y.) {\bf 83}, 491 (1974); B. D. Serot and J. D.
Walecka, Adv. Nucl. Phys. {\bf 16}, 1 (1986);
 Int. Journ. Mod. Phys. {\bf E6}, 515 (1997).

\bibitem{kaplan:1986} D.~B.~Kaplan and A.~E.~Nelson,
 Phys. Lett. {\bf B175}, 57 (1986).

\bibitem{lee:1996} C.~H.~Lee, Phys. Rep. {\bf 275}, 255 (1996)
 and references therein; C. H. Lee, G. E. Brown, D. P. Min, and M. Rho,
 Nucl. Phys. {\bf A585}, 401 (1995).

\bibitem{Li:1997} G. Q. Li, C.-H. Lee, and G. E. Brown,
 Phys. Rev. Lett. {\bf 79}, 5214 (1997).

\bibitem{brown:1998} G. E. Brown, C.-H.Lee, and R. Rapp,
 Nucl. Phys. {\bf A639}, 455c (1998).

\bibitem{muto:2004} T. Muto, Progress of Theoretical Physics
Supplement {\bf 153}, 174 (2004).

\bibitem{thorsson:1994} V. Thorsson, M. Prakash, and J.~M.~Lattimer,
 Nucl. Phys. {\bf A572}, 693(1994); {\bf A574}, 851 (1994), Erratum.

\bibitem{pons:2000} J. A. Pons, S. Reddy, J. Ellis, M. Prakash, and J.
M. Lattimer, Phys. Rev. {\bf C62}, 035803 (2000).

\bibitem{pal:2000} S. Pal, D. Bandyopadhyay, and W. Greiner,
Nucl. Phys. {\bf A674}, 553 (2000).

\bibitem{banik:2001} S. Banik, and D. Bandyopadhyay,
Phys. Rev. {\bf C63}, 035802 (2001).

\bibitem{brown:1988} G. E. Brown, K. Kubodera, D. Page, and P. Pizzochero,
Phys. Rev. {\bf D37}, 2042 (1988).

\bibitem{tatsumi:1988} T. Tatsumi, Progress of Theoretical Physics
{\bf 80}, 22 (1988).

\bibitem{fujii:1994} H. Fujii, T. Muto, T. Tatsumi, and R. Tamagaki,
 Phys. Rev. {\bf C50}, 3140 (1994);
 Nucl. Phys. {\bf A571}, 758 (1994).

\bibitem{kubis:2003} S.~Kubis and M.~Kutschera, Nucl. Phys. {\bf A720},
189 (2003).

\bibitem{pons:2001a} J. A. Pons, AJ. Miralles, M. Prakash, and J.
M. Lattimer, Astrophys. J. {\bf 553}, 382 (2001).


\bibitem{glen_schaf}  N. K. Glendenning and J. Schaffner-Bielich,
 Phys. Rev. Lett. {\bf 81}, 4564 (1998);
 Phys. Rev. {\bf C60}, 025803 (1999).

\bibitem{Zhou:2004} X. R. Zhou, G. F. Burgio, U. Lombardo, H.-J. Schulze, and
W. Zuo, Phys. Rev. {\bf C69}, 018801 (2004).

\bibitem{zuo:2004} W. Zuo, A. Li, Z. H. Li, and U. Lombardo,
 Phys. Rev. {\bf C70}, 055802 (2004).

\bibitem{tatsumi:1998} T. Tatsumi, and M. Yasuhira,
 Phys. Lett. B {\bf 441}, 9 (1998);
 Nucl. Phys. {\bf A670}, 218c (2000).

\bibitem{muto:2000} T. Muto, T. Tatsumi, and N. Ywamoto,
 Phys. Rev. D {\bf 61}, 063001 (2000); {\bf 61}, 083002 (2000).

\bibitem{li:2006} A. Li, G. F. Burgio, U. Lombardo, and
W. Zuo, in preparation.

\bibitem{bombaci:1991} I. Bombaci, and U. Lombardo,
 Phys. Rev. {\bf C44}, 1892 (1991).

\bibitem{zuo:1999} W. Zuo, I. Bombaci, and U. Lombardo,
 Phys. Rev. {\bf C60}, 024605 (1999).

\bibitem{jeukenne:1976}J.P. Jeukenne, A. Lejeune and C. Mahaux,
 Phys. Rep. {\bf 25}, 83 (1976).

\bibitem{song:1998} H.~Q.~Song, M.~Baldo, G.~Giansiracusa, and U.~Lombardo,
 Phys. Rev. Lett. {\bf 81}, 1584 (1998); M. Baldo, A. Fiasconaro, H. Q. Song,
 G.~Giansiracusa, and U.~Lombardo, Phys. Rev. {\bf C65}, 017303 (2002).

\bibitem{wiringa:1995} R. B. Wiringa, V. G. J. Stoks, and R.
Schiavilla, Phys. Rev. {\bf C51}, 38 (1995).

\bibitem{baldo:1997} M.~Baldo, I.~Bombaci, and G.~F. Burgio,
Astron. Astrophys. {\bf 328}, 274 (1997).

\bibitem{pudliner:1995} B.~S.~Pudliner, V.~R.Pandharipande, J.~Carlson, and
R.~B.~Wiringa, Phys.~Rev.~Lett. {\bf 74}, 4396 (1995);
B.~S.~Pudliner, V.~R.Pandharipande, J.~Carlson, S.~C.~Pieper, and
R.~B.~Wiringa, Phys.~Rev. {\bf C56}, 1720 (1997).


\bibitem{grange:1989} P. Grang\'{e}, A. Lejeune, M. Martzolff, and
 J.-F. Mathiot, Phys. Rev. {\bf C40}, 1040 (1989).


\bibitem{zuo:2000} A. Lejeune, U. Lombardo, and W. Zuo, Phys. Lett.
 {\bf B477}, 45 (2000); W. Zuo, A. Lejeune, U. Lombardo, and
  J.-F. Mathiot, Eur. Phys. J {\bf A14}, 469 (2002).

\bibitem{Shapiro:1983} S. L. Shapiro, and S. A. Teukosky, {\it
Black Holes, White Dwarfs, and Neutron Stars}, (John Wiley, New
York, 1983).

\bibitem{brown:1992} G.~E.~Brown, K.~Kubodera, M.~Rho, and V.~Thorsson,
 Phys. Lett. {\bf B291}, 355 (1992).

\bibitem{politzer:1991} H. D. Politzer and M. B. Wise,
 Phys. Lett. {\bf B273}, 156 (1991).

\bibitem{donoghue:1985}
J.~F.~Donoghue and C.~R.~Nappi, Phys. Lett. {\bf B168}, 105 (1986).

\bibitem{dong:1996} S. J. Dong, J.-F. Laga\"{e}, K. F. Liu,
Phys. Rev. {\bf D54}, 5496 (1996).

\bibitem{baym} G. Baym , Phys. Rev. Lett. {\bf 30}, 1340 (1973).


\bibitem{Tolman:1939} R. C. Tolman, Phys. Rev. {\bf 55}, 364 (1939).

\bibitem{Oppenheimer:1939} J. R. Oppenheimer and G. M. Volkoff,
Phys. Rev. {\bf 55}, 374 (1939).


\bibitem{negele:1973} J. W. Negele, and D. Vautherin, Nucl. Phys.
{\bf A207}, 298 (1973).

\bibitem{feynman:1949} R. Feynman, F. Metropolis, and E. Teller,
Phys. Rev. {\bf 75}, 1561 (1949).

\bibitem{baym:1971} G. Baym, C. Pethick, and D. Sutherland,
Astrophys. J. {\bf 170}, 299 (1971).


\bibitem{knor:1995} R. Knorren, M. Prakash, and P. J. Ellis,
Phys. Rev. {\bf C52}, 3470(1995).

\bibitem{schaf:1996} J. Schaffner, and I. N. Mishustin,
Phys. Rev. {\bf C53}, 1416(1996).


\bibitem{mi:1979} A. B. Migdal, in: Mesons in Nuclei, vol.3, eds. M. Rho
and D. Wilkinson
(North-Holland, Amsterdam, 1979)

\bibitem{glen:1992} N. K. Glendenning, Phys. Rev. {\bf D46}, 1274(1992).


\bibitem{maru:2006} T. Maruyama, T. Tatsumi, D. N. Voskresensky,
T. Tanigawa, T. Endo, and S. Chiba, Phys. Rev. {\bf C73}, 035802
(2006).


\bibitem{bbs} M. Baldo, G. F. Burgio, and H.-J. Schulze,
Phys. Rev. {\bf C58}, 3688(1998); Phys. Rev. {\bf C61},
055801(2000).

\bibitem{vid2000} I. Vida$\rm \tilde{n}$a, A. Polls, A. Ramos, L. Engvik, and
M. Hjorth-Jensen, Phys. Rev. {\bf C62}, 035801(2000).


\bibitem{ell:1995} P. J. Ellis, R. Knorren, and M. Prakash,
Phys. Lett. {\bf B349}, 11(1995).






\end{thebibliography}
\end{document}